\documentclass[a4paper]{jpconf}
\usepackage{graphicx}
\usepackage{lineno} 
\usepackage{wrapfig}

\begin{document}
\title{The First Year IceCube-DeepCore Results}
\author{Chang Hyon Ha$^{1, 2}$ for the IceCube Collaboration}
\address{$^{1}$ Department of Physics, Pennsylvania State University, 
  University Park, PA 16802, USA }
\address{$^{2}$ Lawrence Berkeley National Laboratory, Berkeley, CA 94720, USA}
\ead{chha@lbl.gov}

\begin{abstract}
The IceCube Neutrino Observatory includes a tightly spaced inner array 
in the deepest ice, called DeepCore, which gives access 
to low-energy neutrinos with a sizable surrounding 
cosmic ray muon veto. 
Designed to be sensitive to neutrinos at energies 
as low as 10~GeV, DeepCore will be used to study 
diverse physics topics with neutrino signatures, such as
dark matter annihilations and atmospheric neutrino oscillations.
The first year of DeepCore physics data-taking has been completed, 
and the first
observation of atmospheric neutrino-induced cascades with IceCube and DeepCore are presented. 
\end{abstract}

\section{Introduction}
The DeepCore extension to IceCube, shown in Fig. \ref{geo},
triggers on
atmospheric neutrinos at energies 
between about 10~GeV and 1~TeV \cite{dcpaper, chpaper}.
The understanding of the production and oscillations of 
the neutrinos at these energies is intrinsically interesting \cite{osc},
not least because these neutrinos constitute an important background
to astrophysical signal searches,
such as neutrinos from WIMP annihilations \cite{dm}
and neutrinos from soft-spectrum point sources \cite{dcable}.

To observe neutrinos in this energy range, 
DeepCore relies on compact sensor spacing, 
high quantum efficiency photomultiplier tubes (PMTs), 
deployment in the clearest ice,
and a lower trigger threshold than 
the surrounding IceCube detector \cite{dcpaper}.
We report the DeepCore performance with 79 strings of IceCube operating (IC-79),
and highlight results from the first observation of atmospheric 
neutrino-induced cascades in IceCube.

\section{IC-79 Data}
The IC-79 data collected between May 31, 2010 and May 13, 2011 
have been processed and analyzed.
The raw data include a series of waveforms read out 
from digital optical modules (DOMs) in two modes \cite{dom}.
In hard local coincidence (HLC) mode, 
in which both the primary DOM and 
the nearest or next nearest neighbor DOM
report a hit within a $\pm$1000~ns time window,
full waveform digitization is acquired.
If a DOM is in soft local coincidence (SLC) mode 
without neighboring hits,
a reduced waveform data is acquired
consisting of the three digitization bins from the first 16 samples: 
the highest bin and its two neighboring bins.
Therefore, hits in this paper mean DOM readouts 
in HLC plus SLC modes unless specified.
Software is used to remove
spatially and temporally isolated
SLC hits due to noise.
The additional SLC hit information
improves event reconstruction, 
background rejection, and particle identification especially for low multiplicity events.

\begin{wrapfigure}{r}{0.5\textwidth}
  \begin{center}
    \includegraphics[width=0.46 \textwidth]{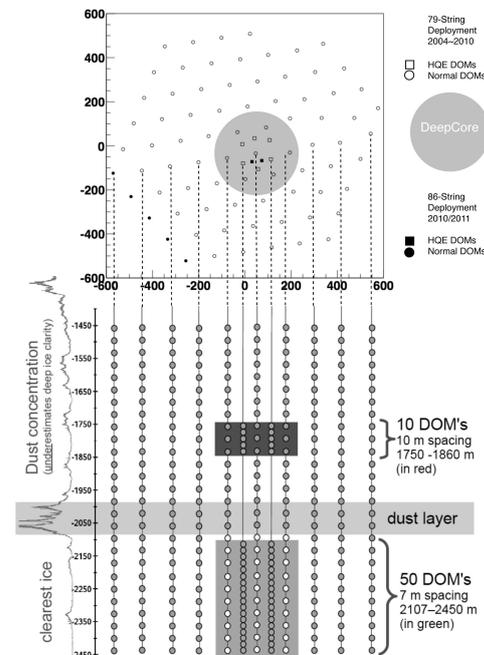}
  \end{center}
  \caption{Schematic top and side views of IceCube 
    in its 79-string configuration. 
    The DeepCore sub-array is defined 
    with 6 densely deployed strings and the 7 nearby standard strings. 
    The relative dust concentration and 
    the DeepCore fiducial volume below the dust layer are also shown.}
  \label{geo}
  \vspace{-1pt}
\end{wrapfigure}

A low threshold trigger (SMT3), demanding 3 or more HLC hits 
within a time window of 2500~ns, is applied to DOMs in the fiducial region 
(the shaded area below the dust band in Fig. \ref{geo}).
Additionally, upgraded PMTs in the DeepCore strings with a $\sim$35\% increase 
in quantum efficiency compared to standard IceCube PMTs, 
help trigger on neutrinos with energies as low as 10~GeV \cite{hqpmt}.

\section{Observation of Atmospheric Neutrino-induced Cascades}
Atmospheric neutrinos are the decay products of 
charged mesons ($\pi^{\pm}$, $K^{\pm}$) 
produced in cosmic ray collisions with nucleons in the atmosphere. 
Cascades are produced 
by charged-current (CC) electron and tau neutrino interactions, 
and neutral-current (NC) neutrino interactions of any flavor,
and create spherically-symmetric light distributions in ice.
Although many atmospheric neutrino-induced muons, long tracks 
created by CC interactions, 
have been collected by IceCube \cite{lab2},
cascades have not been conclusively observed in other IceCube analyses \cite{chpaper, cascade1, cascade2} due to lack of a sufficient veto against the cosmic ray muons, and their low rate \cite{flux}.

\subsection{Background}
The dominant backgrounds for the atmospheric cascades consist of  
cosmic ray muons that mimic signal events
and $\nu_{\mu}^{\rm CC}$ events with dim tracks.
In conventional atmospheric $\nu_{\mu}^{\rm CC}$ detection, 
muon direction information is used to reject background
while a cascade analysis identifies the light pattern of the showers
and enforces containment of the signal.
Therefore, veto techniques with strict signal containment in DeepCore
were developed to remove more than six orders of magnitude 
of background events while retaining reasonable signal efficiency 
for atmospheric neutrino-induced cascades in the fiducial volume. 
\subsection{Event Selection}
IC-79 collected data for 348 days. Over 90\% of the data are high quality
and are used for physics analyses.
A DeepCore on-line filter
is run on the SMT3 triggered event sample at the South Pole.
The pass rate is 17.5~Hz.
The filtered data are sent north for subsequent processing.
The filter algorithm \cite{dcpaper} starts 
by calculating the center of gravity (COG)
of all HLC hits in the fiducial volume 
to get an interaction vertex and time estimate.
Then, the filter disregards any events consistent with 
a cosmic ray muon entering the detector volume by examining the speed
between an individual HLC hit in the veto volume and the COG.
A factor of 10 reduction in data compared to the triggered events 
are reached by this algorithm 
while keeping 99\% atmospheric neutrinos that interact in the fiducial volume.
After applying noise cleaning algorithms that remove hits which 
are not correlated in space and time with other hits,
events with at least 8 total remaining hits
and at least 4 hits in the fiducial region are selected.

The next background rejection is performed 
by a machine learning technique 
with a Boosted Decision Tree (BDT) \cite{bdt}.
The ``BDT5'' is formed from five observables
which are constructed from the spherical hit pattern, 
localized time structure, and quick charge deposition, identifying 
cascade shapes in the detector.
The BDT5 selection reduces the data rate to 0.1~Hz,
achieving a factor of almost 1800 reduction with respect to the trigger.
The atmospheric $\nu_{e}$ rate is 
predicted to be $6.2 \times 10^{-4}$~Hz (about 180 times lower than data rate), 
corresponding to 63\% retention with respect to the trigger.

The reduced data set was then processed 
with iterative likelihood reconstructions 
taking into account detailed Cherenkov 
light propagation in the ice \cite{amareco}. 
Hits falling outside the time window [-3000~ns, +2000~ns] 
with respect to the trigger, or outside of a 150~m radius of a neighboring 
hit within a 750~ns time window, are removed. 
Then, with the remaining hits, after demanding 8 or more hits 
in the fiducial region within a 1000~ns sliding window, 
another BDT is formed with 7 input parameters (BDT7).
Two variables measure the locations of the earliest hits 
in terms of radial and vertical coordinates to select contained events. 
The next three variables separate cascade-like events from muon-like events;
an event is split in half and then charge deposition, COG, 
and particle speed are compared between the two separate halves.
The remaining two variables compare likelihoods of a cascade hypothesis to 
that of a muon hypothesis.
The BDT7 selection reduces the atmospheric muon background 
to $5.0 \times 10^{-4}$ Hz by rejecting a factor 
of 200 more ($3.6 \times 10^5$ cumulatively) 
background while retaining $\sim$40\% of the $\nu_e$ 
signal ($2.6 \times 10^{-4}$ Hz) compared to the previous BDT5 cut.

\subsection{Results}
\begin{figure}[!t]
  \vspace{5mm}
  \centering
  \includegraphics[width=3.1in]{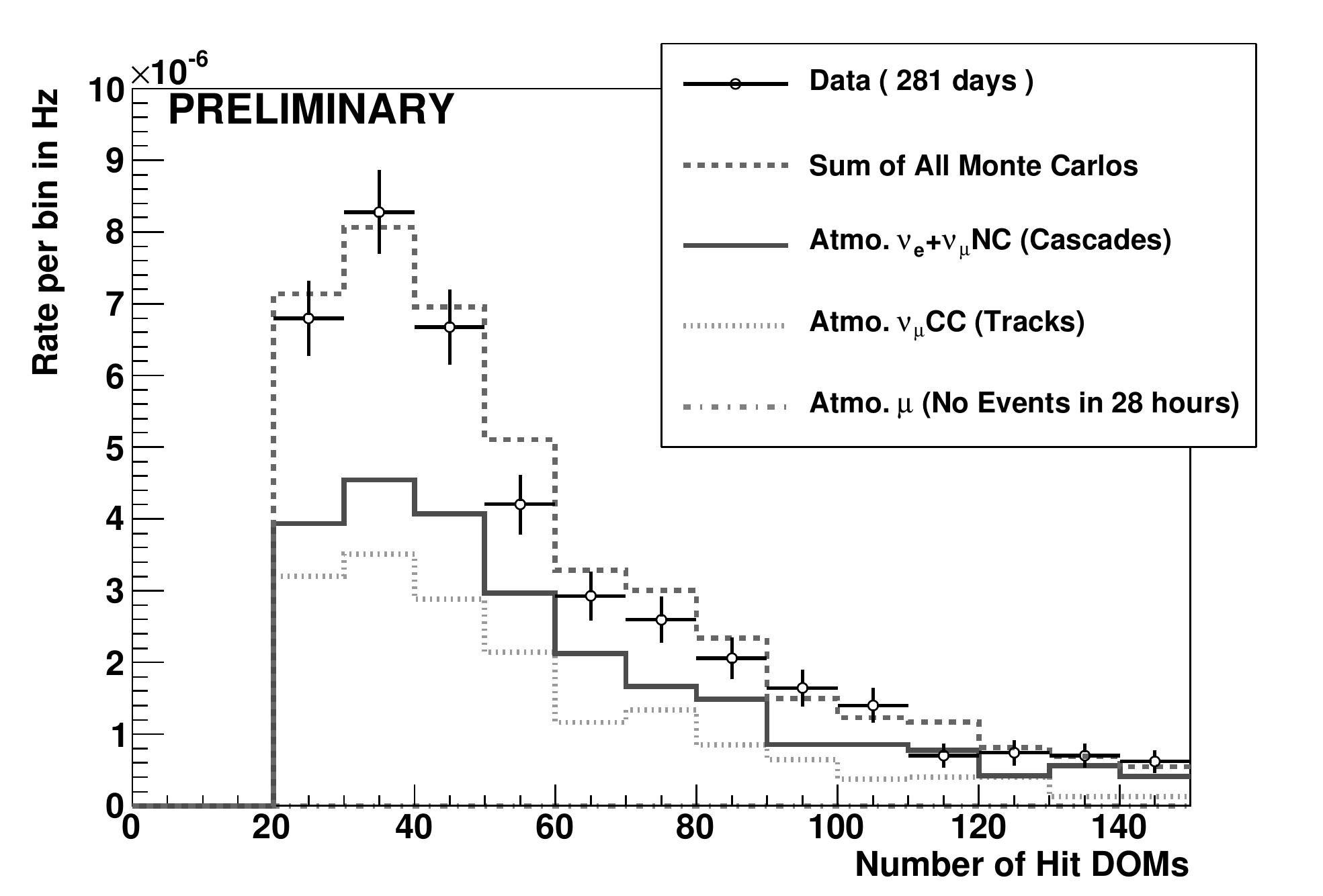}
  \includegraphics[width=3.1in]{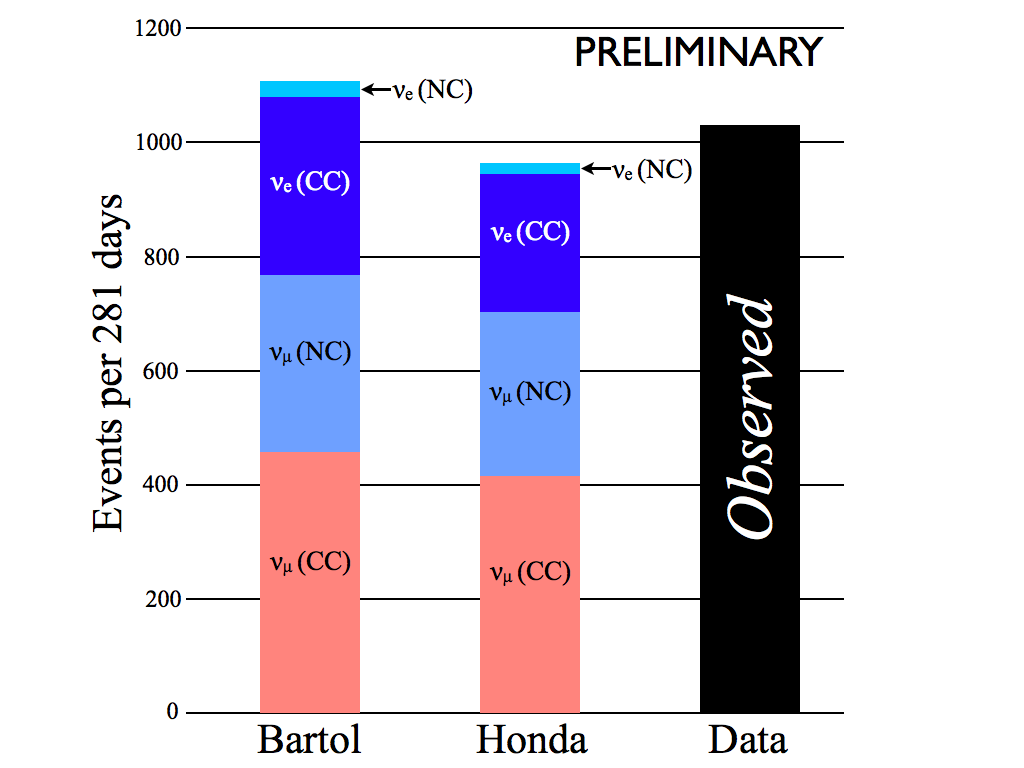}
  \caption{The left plot shows the event rate as a function 
    of the number of hit DOMs. The sum of all Monte Carlo samples is consistent 
    with 281 days of data rate. The cascades are expected to contribute 
    59\% and the tracks are expected to contribute 41\%. No atmospheric muon 
    background events are left in 28 hours of simulated data.
    The oscillation effect is less than 3\% at these energies and is neglected here.
    The bar histogram on the right indicates MC contributions from different
    interactions with two atmospheric flux models (Bartol \cite{bartol} 
    and Honda \cite{honda})
    and the observed data rate. Errors are statistical only.
  }
  \label{hard}
\end{figure}
A set of tight cuts is made on the previously selected BDT7 sample which contains
a large fraction of atmospheric neutrinos.
The cuts aim for high purity cascade detection by rejecting
as many $\nu_{\mu}^{\rm CC}$ events as possible.
Containment cuts based on the vertex depth measurements 
ensure that most signal events are well contained 
inside the DeepCore fiducial volume. 
They select a volume smaller than 
the nominal DeepCore fiducial volume to identify an 
outgoing track from a $\nu_{\mu}^{\rm CC}$ interaction.
Reconstruction quality cuts select events that fit a cascade hypothesis better, 
as measured by the log likelihood from a fit.
Additionally, the selection includes a stronger cut on the BDT7 parameter and 
a requirement of $\ge$20 hit DOMs
to remove the remaining cosmic ray muon background.
As shown in Figure \ref{hard}, with 281 days of data, we observe 1029 total events 
and expect 651 cascades and 455 tracks from simulation using the
Bartol atmospheric neutrino flux model \cite{bartol}.
A lower rate prediction from the Honda model \cite{honda} 
due to a different treatment 
of kaon production in the atmosphere \cite{flux, kaon2}
is presented in the right histogram of Fig. \ref{hard} 
and in Table \ref{table_results}.
The remaining simulated $\nu_{\mu}^{\rm CC}$ events have short muons 
with a median track length of 80~m
where the muon tracks are not detected by this analysis.
About 50\% of the cascades are predicted to be $\nu_{e}$ events and
the balance $\nu_{\mu}^{\rm NC}$ events.
The mean cascade energy is 180~GeV, high enough 
that $\nu_{\mu} \rightarrow \nu_{x}$ oscillations has a small ($<$3\%) effect. 
The atmospheric muon simulation predicts zero events in 28~hours. 
Systematic errors are not included.

\section{Conclusion}
We report on the first observation of atmospheric neutrino-induced cascade events 
with IceCube and DeepCore.
The preliminary summary of the results is shown in Table~\ref{table_results}. 
Systematic errors originating from ice modeling, 
detection efficiency of DOMs, neutrino-nucleon cross-sections, 
and atmospheric neutrino flux model are under evaluation.
In the near future, data analyses using a similar technique 
as that presented here, focusing
on neutrino oscillations, WIMP searches, and 
neutrino surveys for the southern sky, are expected.

\begin{table}[t]
\begin{center}
\begin{tabular}{c|ccc||c|c|c}
\hline\hline
&\multicolumn{3}{|c||}{$\rm C^{sig}$} & $\rm C^{bg}$ & MC Sum & $\rm N^{obs}$\\
Type&$\nu_{e}^{\rm NC}$ & $\nu_{e}^{\rm CC}$ & $\nu_{\mu}^{\rm NC}$ & $\nu_{\mu}^{\rm CC}$ & &\\
\hline\hline
Bartol & 25 & 312 & 314 & 455 & 1106 &-\\
\hline
Honda & 18 & 245 & 287 & 415 & 965 &- \\
\hline
Data & - &- &-  & - & - & 1029\\
\hline
\end{tabular}
\caption{The number of events are shown with final selections. 
  $\rm N^{obs}$ means observed events in 281 days of real data. 
  $\rm C^{sig}$ and $\rm C^{bg}$ refer predictions 
  of the cascade signal and its background respectively.
  The MC numbers use 281 days normalization and their statistical errors 
  are $\sim$3\%.
}
\label{table_results}
\end{center}
\end{table}


\section*{References}
\begin{thebibliography}{9}
\bibitem{dcpaper} R. Abbasi \textit{et al.} [IceCube Collaboration] arXiv:1109.6096.
\bibitem{osc} R. Abbasi \textit{et al.} [IceCube Collaboration] arXiv:1111.2731.
\bibitem{chpaper} R. Abbasi \textit{et al.} [IceCube Collaboration] arXiv:1111.2736.
\bibitem{dm} R. Abbasi \textit{et al.} [IceCube Collaboration] arXiv:1111.2738.
\bibitem{dcable} C. Wiebusch [IceCube Collaboration] arXiv:0907.2263.
\bibitem{dom} R. Abbasi \textit{et al.} [IceCube Collaboration], Nucl. Instrum. Meth. \textbf{A601}, 294 (2009).
\bibitem{hqpmt} D. J. Koskinen [IceCube Collaboration], WSPC-Proceedings WIN11, (2011).
\bibitem{lab2} R. Abbasi \textit{et al.} [IceCube Collaboration], Phys. Rev. \textbf{D83}, 012001 (2011).
\bibitem{flux} T. K. Gaisser, AIP Conf. Proc., \textbf{944}, 140; arXiv:astro-ph/0612274 (2007).
\bibitem{cascade1} R. Abbasi \textit{et al.} [IceCube Collaboration], Astropart. Phys. \textbf{34}, 420 (2011).
\bibitem{cascade2} R. Abbasi \textit{et al.} [IceCube Collaboration], Phys. Rev. \textbf{D84}, 072001 (2011).
\bibitem{bdt} H. Voss \textit{et al.}, Proc. Sci. ACAT, 040 (2007).
\bibitem{amareco} J.Ahrens \textit{et al.} [AMANDA Collaboration], Nucl. Instrum. Meth. \textbf{A524}, 169 (2004).
\bibitem{bartol} G. D. Barr \textit{et al.}, Phys. Rev. \textbf{D70}, 023006 (2004).
\bibitem{honda} M. Honda \textit{et al.}, Phys. Rev. \textbf{D75}, 043006 (2007).
\bibitem{kaon2} T. K. Gaisser, Nucl. Phys. Proc. Suppl. \textbf{118}, 109 (2003).
\end{thebibliography}
\end{document}